# 3d Transition Metals and Oxides within Carbon Nanotubes by Co-Pyrolysis of Metallocene & Camphor: High Filling Efficiency and Self-Organized Structures

Aakanksha Kapoor[1], Nitesh Singh[1], Arka Bikash Dey[2], A.K. Nigam[3], Ashna Bajpai*[1, 4]

[1]*Division of Physics, Indian Institute of Science Education and Research (IISER), Dr. Homi Bhabha Road, Pune, 411008, India.*

[2]*Saha Institute of Nuclear Physics, 1/AF Bidhannagar, Kolkata, India.*

[3]*Department of Condensed Matter Physics and Material Science, Tata Institute of Fundamental Research, Homi Bhabha Road, Mumbai.*

[4]*Center for Energy Science, Indian Institute of Science Education and Research (IISER), Dr. Homi Bhabha Road, Pune, 411008, India.*

**Abstract**

We demonstrate that a *single zone furnace* with a modified synthesis chamber design is sufficient to obtain metal (Fe, Co or Ni) filled carbon nanotubes (CNT) with high filling efficiency and controlled morphology. Samples are formed by *pyrolysis of metallocenes,* a synthesis technique that otherwise requires a *dual zone furnace*. Respective metallocene in all three cases are sublimed in powder form, a crucial factor for obtaining high filling efficiency. While Fe@CNT is routinely produced using this technique, well-formed Ni@CNT or Co@CNT samples are reported for the first time. This is achieved by sublimation of nickelocene (or cobaltocene) in combination with 'camphor'. These samples exhibit some of the highest saturation magnetization ($M_s$) values, at least an order of magnitude higher than that reported for Ni or Co@CNT, by aerosol assisted pyrolysis. The results also elucidate on why Ni or Co@CNT are relatively difficult to obtain by pyrolyzing powder metallocene alone. Overall, a systematic variation of synthesis parameters provides insights for obtaining narrow length and diameter distribution and reduced residue particles outside filled CNT - factors which are important for device related applications. Finally, the utility of this technique is demonstrated by obtaining highly aligned forest of $Fe_2O_3$@CNT, wherein $Fe_2O_3$ is a functional magnetic oxide relevant to spintronics and battery applications.

*Corresponding author: Tel +91-20-2590-8107. E mail: ashna@iiserpune.ac.in

# 1. Introduction

Nanoscale magnetic materials, when encapsulated within the CNT are multifunctional hybrids, important from both fundamental and application point of view. While CNT are well known to possess exceptional electrical, thermal and mechanical properties [1], the magnetic encapsulate can give rise to a broad diversity in both magnetic and electrical properties, depending on its type, size and morphology [2]. Hitherto, the most popular magnetic encapsulate has been Fe, which is a ferromagnetic (FM) metal in bulk. When confined within CNT, the Fe@CNT hybrids can have applications in areas ranging from medical science, energy sector and spintronics [3-8]. The encapsulation of transition metals and metal oxides can bring forward new possibilities as metal oxides are well known for a host of novel electronic and magnetic ground states [2, 9]. For instance, α- $Fe_2O_3$ is an AFM insulator and a piezomagnet [10], whereas $Fe_3O_4$ is an FM (half) metal that possesses spin polarization larger than Fe metal. The oxides of other transition metals, such as chrome oxides can be half metals or magneto electric in their bulk state and exhibit remarkable emergent properties upon nano-scaling [11-20]. These metals and oxides, when confined and protected within CNT, can thus provide huge tunability in their functional properties as well as give rise to new emergent phenomena that are possible at the interface of these hybrids [9, 13-16].

We also emphasize that an obvious choice is to first encapsulate first few 3d transition metals, out of which most reports are on Fe@CNT [21-27]. The encapsulation of other metallic FM such as Co or Ni, as well as a wide variety of their respective oxides, is rarely explored [18-20, 28-30]. Investigating these hybrids in the form of either metal, or metal oxide encapsulated within the core cavity of the CNT demands: (i) significant filling of magnetic encapsulate within the core cavity, (ii) better control of diameter and length of the filled CNT and, (iii) cost-effectiveness and reproducibility of the synthesis procedure. All these three points are important, keeping in view the potential device applications and the fundamental aspects related to a nano scale multifunctional magnetic material within a one-dimensional conductor [2, 9].

Organometallic compounds, such as *ferrocene*, are routinely used as a precursor for the catalytic synthesis of carbon nanotubes [19-27, 32]. On a more general note, *ferrocene* belongs to the family of *metallocene* with a formula $M(C_5H_5)_2$, where M can be Fe, Co or Ni. These compounds can provide both the carbon source and the metallic catalyst particles, for facilitating the growth of the CNT. The relatively low sublimation temperature (150-300 °C) makes metallocene the most popular choice as a precursor [20-27]. There are two major routes of using metallocene to form CNT. The first is aerosol assisted/solution based Chemical Vapor deposition (CVD), in which of metallocene is first dissolved in a suitable *liquid* precursor and pyrolyzed [19, 29-36]. The second procedure involves direct sublimation of metallocene in *powder* form [20-27].

Prior literature suggests that aerosol assisted CVD is primarily for the catalytic growth of the pristine CNT, which, in principle, is possible by using any of the three metallocenes [29-36]. The most popular precursor, among these, is still ferrocene. Due to relatively less density of catalytic particles, aligned growth of CNT is also easily achieved, following this procedure [33-35]. Here, an important issue, relevant to the magnetic state of the CNT by aerosol assisted CVD, is that metal catalyst particles which facilitate the growth of CNT, are not only trapped within its core cavity but also adhere outside the CNT. Due to the inherent ferromagnetic nature of these nano particles, the CNT thus formed exhibit ferromagnetic hysteresis in bulk magnetization measurements [36]. The fraction of such residue nano particles is typically small in aerosol assisted CVD, leading to the observed saturation magnetization, Ms ~ 2-10 emu/g in the case of Fe@CNT [31, 36], and 0.1-0.8 emu/g in case of Co or Ni@CNT [19, 30-31]. It is important to note that in the case where only pristine and *non-magnetic* CNT is required, the *as -prepared* CNT are subjected to specific annealing treatments, that lead to the significant reduction of residue nano particles and therefore ferromagnetic traits are seen to disappear [37].

For the formation of CNT along with deliberate (and efficient) filling of ferromagnetic metal within its core cavity, sublimation of metallocene in *powder form* is best suited [20-27]. Here the metal particles not only facilitate CNT growth but also assembles in the form of long nano wires well within the core of the CNT, giving rise to significantly higher filling factor, and consequently substantially larger $M_s$. For instance, $M_s \sim$ 20-60 emu/g has been reported in the case of Fe@CNT [26,27]. It is important to note that the *metal nano particles* adhering outside the CNT can still be a source of error, as their magnetic contribution is difficult to disentangle from the *metal encapsulate* within the core cavity in bulk magnetization measurements.

It is also noteworthy that using solid precursor route, Fe@CNT are easily formed in which filling efficiency (and consequently $M_s$) can be tuned by variations in synthesis parameters [26, 38-39]. However, to the best of our knowledge, there are no reports on Ni and Co@CNT with high filling efficiency by tuning of the synthesis parameters [26] following the same route, even though, the sublimation temperatures of all the three metallocenes are in close vicinity. Ni and Co@CNT are reported either using aerosol assisted CVD, for which $M_s \sim$ 0.1-0.4 emu/g [19, 30-31], or by using the nano-particles grown on substrates [18]. Thus, following the CVD of metallocene in *powder form*, the challenge still is to have: (i) magnetic nano material protected well within the core cavity of the CNT, (ii) minimum residue particles outside the CNT, (iii) narrow length/diameter distribution of Filled CNT, and most importantly, (iv) reproducibility and cost-effectiveness of the synthesis procedure.

In addition to the above mentioned factors, yet another important issue is the morphology of the metal-filled CNT. This can be broadly classified into two types. The first is randomly oriented CNT entangled with each other. This can be referred to as *spaghetti*-like morphology. The second is vertically aligned/self-organized CNT structures, also referred to as *aligned forest* CNT. The filling efficiency, as gauged from $M_s$, can be quite similar. However, the self-organized CNT carpets lead to naturally narrow length and diameter distribution, which is highly desirable for certain nano-electronics/optics/spintronic related applications. It is also important to note that it is easier to obtain vertically aligned CNT following aerosol assisted CVD [33-36]. However, the solid precursor CVD usually leads to a mixture of entangled CNT along with *aligned forests* [26, 27] especially when CNT are deposited directly on the quartz tube reactor. This is due to the fact that higher density of the catalytic nano-particles, which is crucial for high filling efficiency, also promotes entangled CNT growth.

In this work, our focus is in improvisations in CVD using metallocene to obtain Fe, Ni and Co filled CNT with higher filling efficiency and controlled morphology. The structure of the paper is as follows. We first present the details of the experimental set up which involves CVD of metallocene in **powder form**, using a **single zone furnace**. We further demonstrate that all three types of metal filled CNT with significant filling efficiency can be formed using either powder metallocene or the respective metallocene is co-pyrolyzed with a cost effective and environmental friendly compound 'camphor'. It is to be recalled that pristine CNT can be formed by sublimation of camphor alone on suitable substrates [40-41]. However, to the best of our knowledge, this is the first report on obtaining self-organized vertically aligned CNT with significant filling efficiency using metallocene in combination with camphor.

## 2. Experimental

The precursors, high purity Ferrocene, Cobaltocene, Nickelocene (98% pure), and Camphor (96% pure) have been procured from *Sigma-Aldrich*. The Furnace used is from Nabertherm R 100/750/13. The Scanning Electron Microscopy (SEM) images are recorded using ZEISS ULTRA *plus* field-emission SEM. High-resolution Transmission Electron Microscopy (TEM) using FEI-TECNAI microscope at an operating voltage of 200 kV. Raman spectroscopy measurements are done on HORIBA JOBIN YVON LabRAM HR 800 with an excitation wavelength of 488 nm. All the samples have been characterized using X-ray powder diffraction (XRD) using Bruker D8 advance with Cu $K\alpha$ radiation ($\lambda$ = 1.54056 Å). The thermal analysis was determined by a thermogravimetric analyser, PerkinElmer STA 6000, under air at 20 mL/min, at a heating rate of 10 °C min$^{-1}$ from 30 °C to 900 °C. Temperature variation of synchrotron XRD from 20 K-300 K has been conducted in BL-18 beam line, Photon Factory, Japan. The synchrotron XRD data has been fitted using Rietveld Profile Refinement. Bulk magnetization measurements have been performed using Superconducting Quantum Interface Device (SQUID) magnetometer from Quantum Design.

## 3. Results & Discussion

In the usual CVD synthesis, the metallocene is sublimated within a quartz tube reactor, placed inside the first zone of a *two-zone furnace,* at a fixed temperature, referred to as sublimation temperature "$T_{sub}$." The plumes are carried by an inert carrier gas, like Ar, to the pyrolysis (the second) zone of the furnace, held at relatively higher temperature, referred to as the pyrolysis temperature "$T_{pyro}$.". A dual zone furnace with two individual temperature controllers is therefore required to individually control $T_{sub}$ or $T_{pyro}$. For a particular run, $T_{pyro}$ and $T_{sub}$ along with "Ar-flow" rate are the three major variables. These parameters in different combinations are known to give variations in dimension, filling efficiency and morphology of the filled CNT [26, 38-39].

## 3.1: Single Zone Furnace with one Temperature Control Unit

Here we have used a single zone furnace for the solid source CVD set up. The set-up is shown schematically in **Figure 1(a)**. The temperature profile of the furnace is plotted in **Figure 1(b),** which displays a constant temperature zone in the middle region and two variable temperature zones on the either side. The temperature profile is obtained for three different set-temperatures: 700 ºC, 800 ºC and 900 ºC, as CNT are known to form at any of these (fixed) set temperatures [20, 26, 41]. The temperature stability of the middle zone is ±2 ºC and it is marked as $T_{pyro}$ in **Figure 1(b)**. The design of the synthesis chamber and modifications that lead to a better control in L, D, filling efficiency and morphology are discussed in terms of each relevant synthesis parameters.

### *3.1.1: Design of the Synthesis Chamber and Sample Insertion Arrangements:*

The synthesis chamber consists of two coaxial quartz tubes inserted within the *single zone-furnace*. One end of the outer quartz tube is connected to a standard compression seal arrangement for metallocene insertion. An acetone trap is connected at the other end as depicted in **Figure 1(a)**. The compression seal enables precise movement of a quartz rod within the inner quartz reactor, without disturbing the Ar flow rate. An open quartz boat, about 5 cm in length, is fused at the end of this quartz rod. The metallocene in powder (or pellet) form is placed on this boat. This assembly can be precisely moved to any predetermined spot within the inner quartz tube. The inner tube is ~ 1 meter in length and 3 cm in diameter and it serves as the main quartz reactor for the deposition of the CNT.

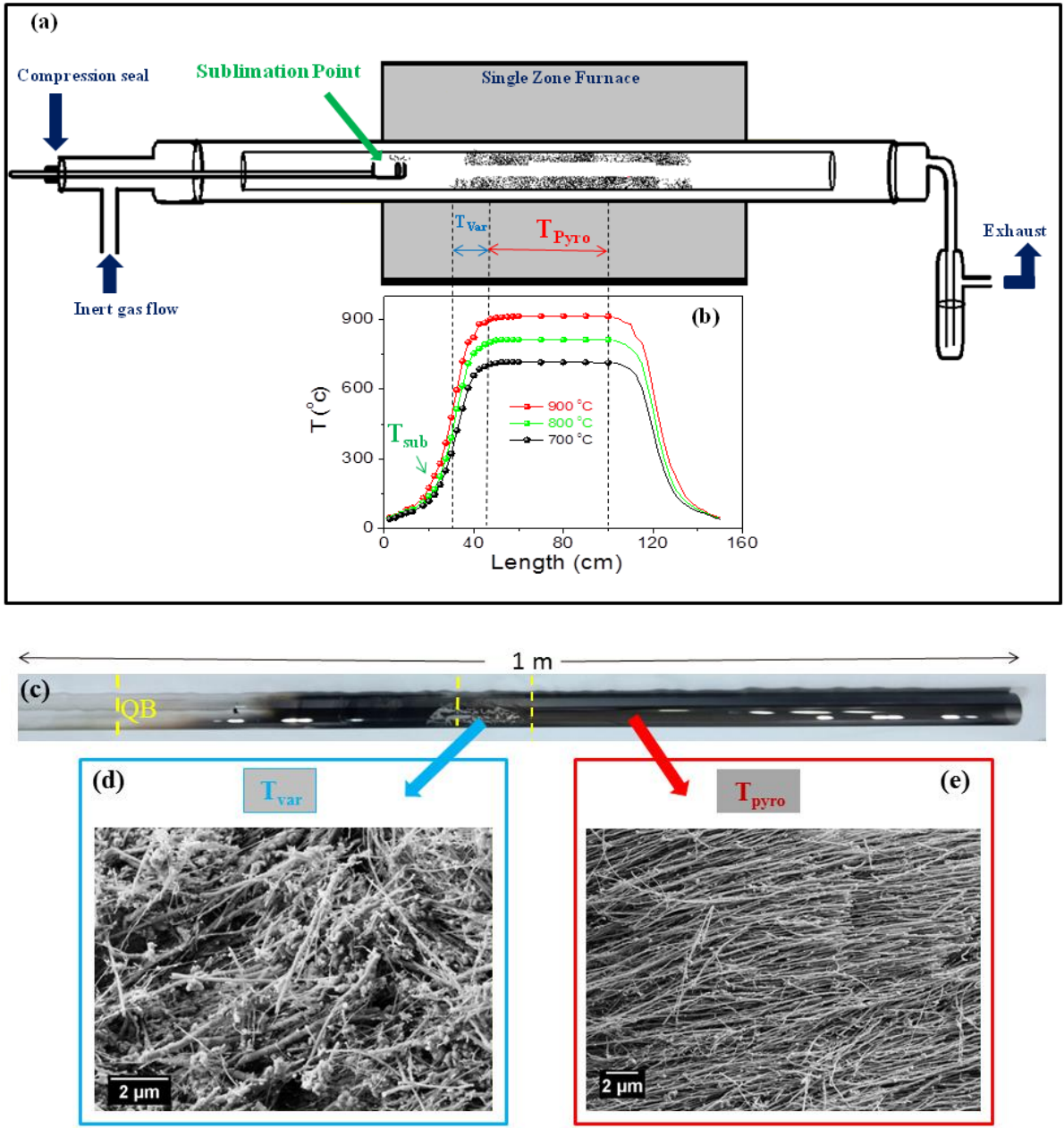

*Figure 1:* *(a) shows a schematic of the single-zone furnace along with the synthesis chamber consisting of two co-axial quartz tubes. A compression seal arrangement enables the movement of a quartz rod fused with a quartz boat in the synthesis chamber. The temperature profile of the furnace at three set temperatures, 900, 800, and 700 °C, is shown in (b). (c) s*hows *a typical inner quartz tube, post synthesis. Here QB signifies position of the quartz boat. (d) & (e) are representative SEM images of Fe@CNT collected from a variable temperature region ($T_{var}$) and actual pyrolysis region ($T_{pyro}$) shown schematically in (a).*

For all the synthesis runs discussed here, the furnace is first set at a fixed temperature, $T_{pyro}$, under continuous Ar-flow. The quartz boat containing the metallocene is kept inside the inner tube at room temperature, till the pyrolysis zone attains the set temperature, within an accuracy of ± 2ºC. Once $T_{pyro}$ is stabilized, the quartz boat is pushed to the sublimation point, shown schematically in **Figure 1(a)**. The reaction time is typically 10 minutes, after which the furnace is cooled down to room temperature under constant Ar-Flow.

### 3.1.2: Sublimation Temperature ($T_{sub}$):

All the three metallocenes sublime at any (fixed) temperature between 150-450 °C. For a fixed temperature in the middle region, a point in the variable temperature region, corresponding to a chosen $T_{sub}$ can be identified from temperature profile shown in **Figure 1(b).** The temperature accuracy of $T_{sub}$ is roughly ±20 °C over ~ 2 cm length, which is the length of the quartz boat in which the metallocene is spread. By inserting the metallocene, either in pellet or in powder form, through the compression seal port, at the predetermined spot enables its sublimation. The reason to pelletize the metallocene prior to sublimation was to further narrow down the temperature accuracy of sublimation region to ±5 °C.

### 3.1.3: Ar-Flow Rate:

The sublimed plumes are carried to the pyrolysis region, ideally in the middle zone of the furnace, where the CNT are formed on the walls of inner quartz tube reactor. We find that the *Ar-flow rate* and sublimation temperature, $T_{sub}$, are not altogether independent experimental parameters as will be evident from the data presented in subsequent sections. For a fixed sublimation temperature, different *Ar-flow rates* also results in variations in the morphology of the final product.

### 3.1.3: Pyrolysis Temperature ($T_{pyro}$):

As evident from the x axis of **Figure 1(b)**, the constant temperature zone or the middle zone of this furnace, for any given $T_{pyro}$, is rather long, about ~ 60 cm in length. The temperature accuracy of this middle region is $\Delta T \sim \pm 2$ °C. A longer pyrolysis zone with better temperature accuracy should enable one to obtain CNT with larger yield and fixed morphology, depending on the chosen value of $T_{pyro}$. However, it is important to note that for any fixed $T_{pyro}$ the sublimed plumes necessarily travel from the *sublimation point*, $T_{sub}$, *to* the onset of $T_{pyro}$. It is also well known that CNT can be formed at any fixed temperature ranging from 500-1000 °C [20, 26, 40]. For example, for a fixed $T_{pyro} \sim 900$ °C in the middle zone, CNT not only form in the middle zone but also deposit on that portion of the inner quartz tube, which falls in the variable temperature zone, marked as $T_{var}$ in **Figure 1(a)&1(b)**. In this region, the temperature varies from 600-900 °C, a range known to be conducive for CNT growth. Even though $T_{pyro}$ is fixed for a particular run, when the quartz boat reactor is taken out after the reaction, the CNT are scraped through the entire inner wall of the quartz tube reactor. This leads to the unintentional mixing of CNT formed at variable temperature region, along with those formed at the constant temperature region.

It is also to be noted that the presence of a variable temperature region cannot be avoided either in a routine two zone furnace based CVD or in the present single zone furnace. However, to circumvent this problem, the portion of the inner quartz tube, falling within the variable and constant temperature zone is clearly marked w.r.t. a fixed point. The sample is collected individually from both the regions[*]. **Figure 1(c)** shows a real time picture of a typical inner quartz boat, post reaction. The representative SEM pictures from Fe@CNT sample collected from the $T_{var}$ and $T_{pyro}$ are shown **Figures 1(d) and 1(e)** respectively. The morphology of the CNT clearly shows the wide difference. A mixture of entangled CNT as well as disrupted forest structures and a large number of residual particles outside CNT are found in variable temperature region.

*3.1.5: Camphor as a co–precursor with metallocene:*

It is well known that Fe-filled CNT are easily formed when powder ferrocene sublimed and pyrolyzed. The sample thus formed exhibit larger filling efficiency [23-26]. However, using powder nickelocene is not known to yield Ni filled CNT. In the present set up, no combination of $T_{pyro}$, $T_{sub}$, and Ar flow rate resulted in any good quality filled CNT, when cobaltocene (or nickelocene) is sublimed alone. For a fixed $T_{pyro}$ and $T_{sub}$, and lower Ar-flow rate, it was observed that a large number of metal nano particles are accumulated near the sublimation point. Increasing the Ar-flow rate resulted in a large number of metal nano particles in the pyrolysis range, again with a negligible presence of CNT. However, the overall data indicated that the carbon content, by the time plumes reach the temperatures conducive to CNT growth, is not sufficient. This is unlike the case of Fe@CNT, where either entangled or forest structures are easily obtained over a wide temperature range. Co-pyrolysis of camphor, which can provide an extra source of carbon, with nickelocene (or cobaltocene), however, enables the growth of clean samples of Ni@CNT (or Co@CNT). For the data presented subsequently, the camphor is thoroughly mixed with metallocene using a mortar pestle and the mixture is used, both as a pellet as well as in powder form. For these runs, we optimized $T_{pyro}$, $T_{sub}$, and Ar-flow rate along with additional synthesis parameter, the ratio of camphor: metallocene.

**3.2: Fe, Co and Ni@CNT using the Single Zone Furnace**

In this section, the results of XRD, SEM, TEM, Raman and Magnetization characterization are presented for Fe@CNT, Co@CNT, and Ni@CNT. The synthesis parameters for all three samples are given in **Table 1**. **Figure 2(a-c)** shows XRD pattern for all three samples. The XRD data in all three cases exhibits the graphitic as well as metal peaks corresponding to CNT and the Fe, Ni or Co encapsulate. **Figure 2(d)** shows the Fe@CNT carpet, obtained from $T_{pyro}$ region. These carpets are up to few 100 μ-m in length and consists of aligned forest of CNT with L ~ 20 μm and D ~ 20-40 nm. The diameter distribution is significantly narrowed down as evident from the **Figure 2(d)**. Due to the collection of the sample only from $T_{pyro}$ region, the results are fairly reproducible in terms of morphology, L and D and filling efficiency, when multiple runs are performed, without varying the synthesis parameters. In the case of Co@CNT, large carpet like structures, again a few 100 μ-m in length, are obtained with aligned forest type structures of CNT with L ~ 8-10 μ-m and D ~ 20-40 nm, as shown in **Figure 2(e)**. The residue particle density outside the CNT is larger in this case. While Fe@CNT and Co@CNT samples are collected from $T_{pyro}$ region, the Ni@CNT sample shown in **Figure 2(f)** is collected from the $T_{var}$ region, which exhibits short entangled CNT.

In order to further confirm the nature of the encapsulate, TEM micrographs were recorded on multiple CNT and a few representative pictures are shown in **Figure 3(a-c)**. These high-resolution TEM images confirm the existence of crystalline metal nanowires encapsulated within the core cavity of the CNT for each case. To the best of our knowledge, this is the first report of sublimation of powder nickelocene (or cobaltocene) for obtaining well formed CNT with significant filling using solid state CVD.

| Sample | Sublimation Temperature, $T_{sub}$ (°C) | Pyrolysis Temperature, $T_{pyro}$ (°C) | Rate of Argon flow (sccm) | Amount of respective metallocene (mg) | Amount of Camphor (mg) | Sample Collection region |
|--------|------|------|------|------|------|------|
| Fe@CNT | 300 | 900 | 2500 | 500 | - | $T_{pyro}$ |
| Co@CNT | 200 | 800 | 2500 | 500 | 100 | $T_{pyro}$ |
| Ni@CNT | 350 | 700 | 2100 | 500 | 100 | $T_{var}$ |

*Table 1: Summary of synthesis parameters for the samples shown in Figure 2-4.*

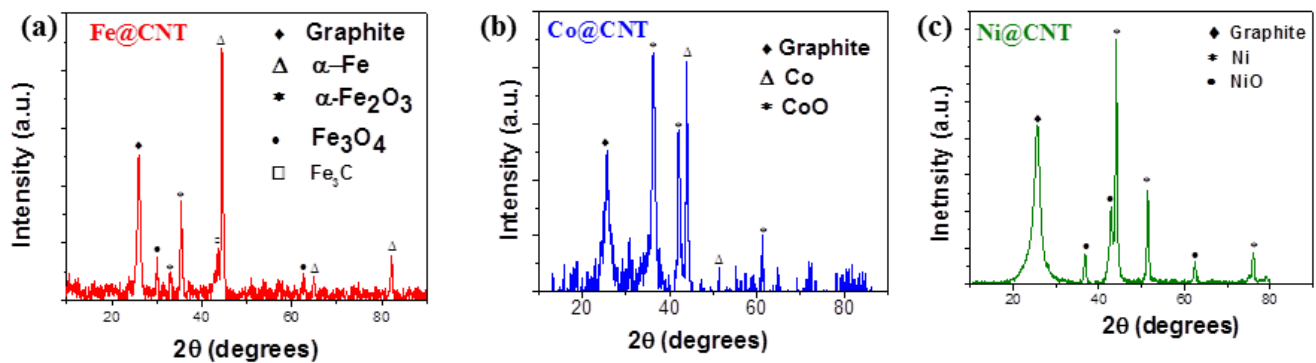
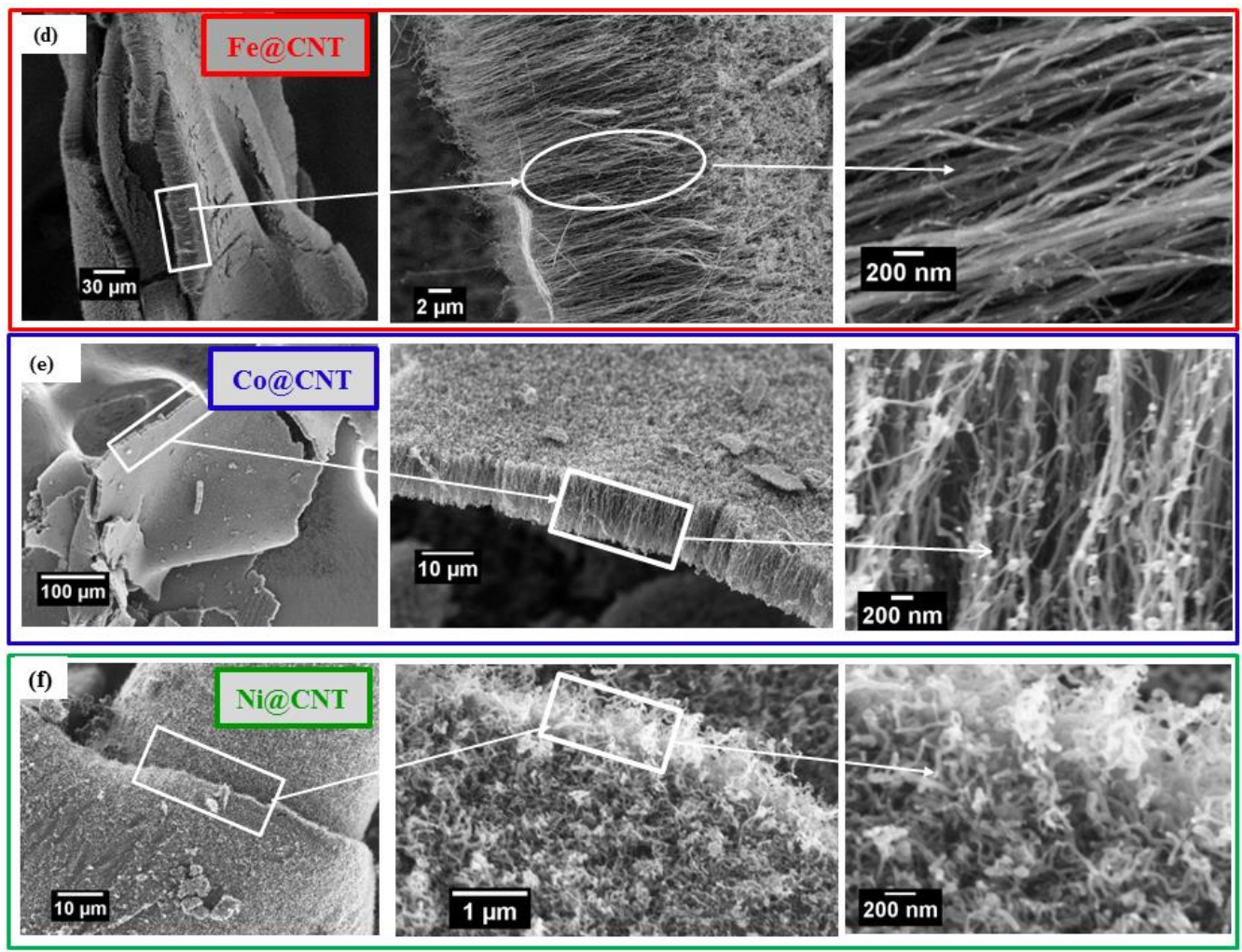

*Figure 2: (a-c) Representative Powder XRD of Fe@CNT, Co@CNT, and Ni@CNT respectively. Representative SEM micrographs of (d) display large carpets of aligned Fe@CNT with arrow L and D distribution. (e) shows the same for Co@CNT aligned forest type structure. (f) Ni@CNT with spaghetti-like structures.*

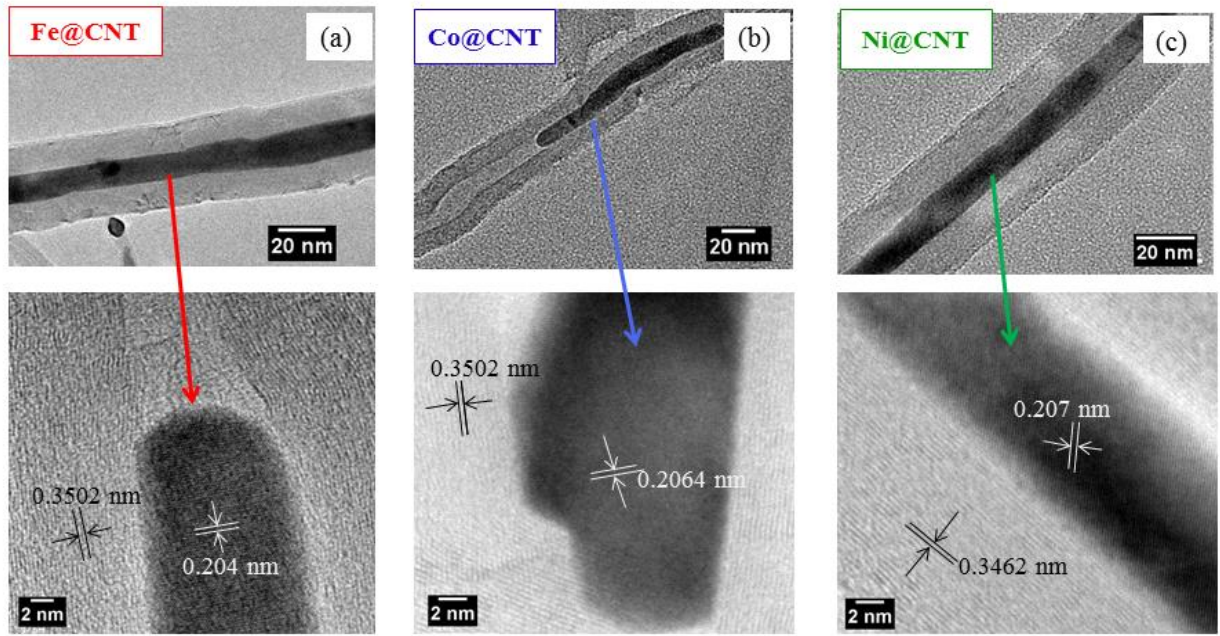

*Figure 3: High-Resolution TEM micrographs for (a)Fe@CNT, (b)Co@CNT, and (c)Ni@CNT. The black and white arrows mark the lattice spacing of the graphitic shells and the encapsulated metal respectively.*

Here a key result is the gain in filling efficiency, especially in the case of Ni and Co@CNT due to the usage of powder metallocene for solid state CVD. This is corroborated by TGA measurements for all three type of samples. The TGA curves obtained under air from 30 °C to 900 °C at a heating rate of 10 °C min$^{-1}$ are shown in Figure 4(a-c). The TGA profiles depict the residual weight % to be ~ 47% for Fe@CNT, 49% for Co@CNT, and 25% for Ni@CNT, confirming the high filling efficiency in these samples. The high filling efficiency is further confirmed from correspondingly high saturation magnetization values, especially in case of Ni and Co@CNT. The M-H isotherms depicting the $M_s$ value for the all three samples are shown in **Figure 4(d-f)**. For Fe@CNT, $M_s$ ~ 40 emu/g, whereas it is ~ 12 emu/g for Ni and Co@CNT. It is to be emphasized that the use of camphor enabled us to obtain Ni and Co filled CNT with significantly improved $M_s$ values (up to 12 emu/g), as compared to what one obtains for aerosol assisted CVD (0.1-0.4 emu/g) reported earlier [18, 30-31]. We reiterate, especially in the case of of Ni@CNT, pyrolysis of only nickelocene using solid state CVD is not known yield well-formed CNT. Though there exist reports of co pyrolysis of nickelocene with ferrocene in aerosol assisted CVD. However, the magnetization data will contain contributions from both Ni and Fe in such cases [30].

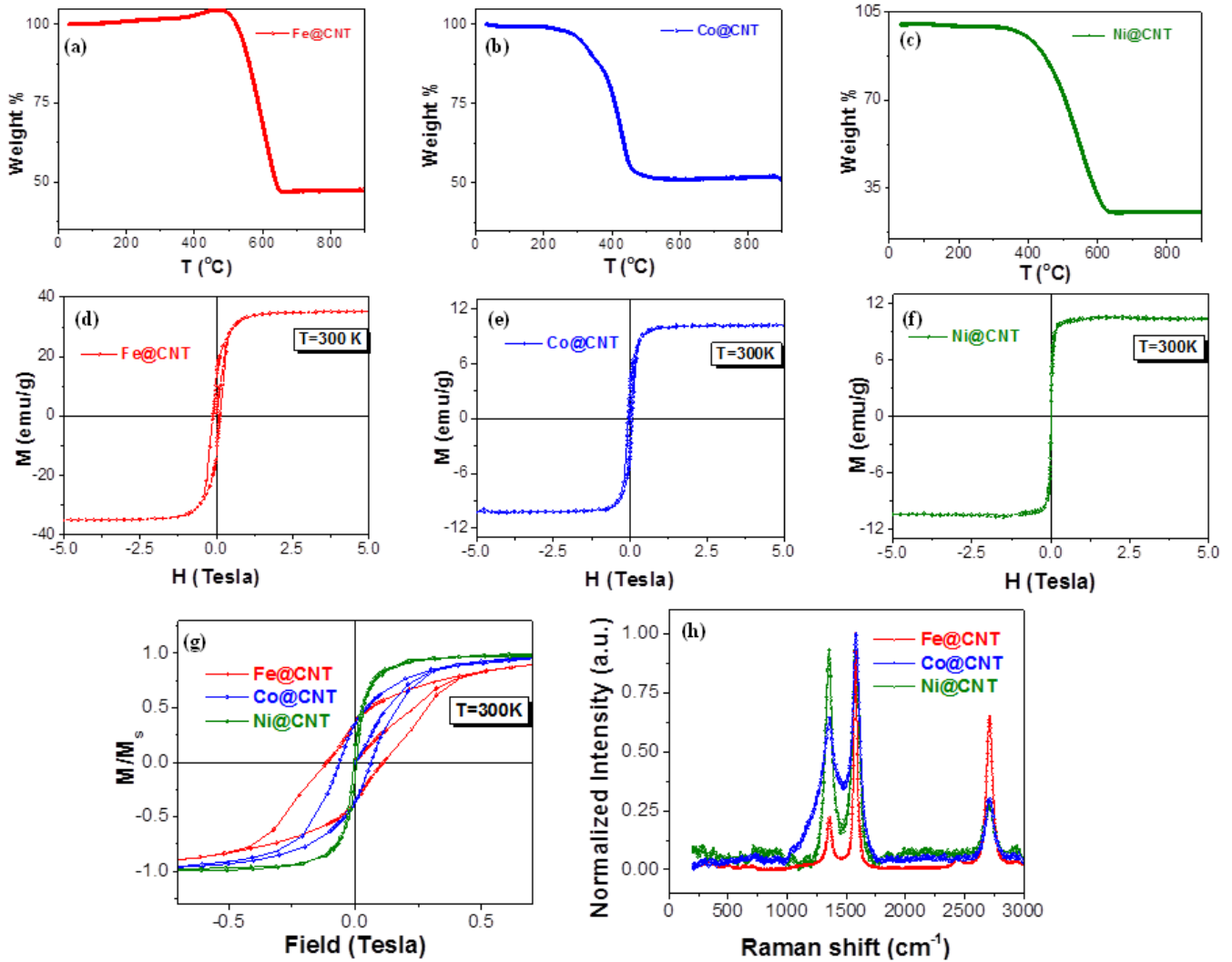

*Figure 4:(a)- (c) display the TGA curves of all the three samples obtained under air from 30 °C to 900 °C at a heating rate of 10 °C min$^{-1}$. The residue weight % was found to be 47% for Fe@CNT, ~ 49% for Co@CNT, and ~25% for Ni@CNT. (d) – (f) display M.H isotherms of all three samples, depicting $M_s$ ~ 40 emu/g in case of Fe@CNT, and 12 emu/g in the case Co@CNT as well as Ni@CNT. (g) Normalized M-H scan at 300 K limited to ±1 kOe, depicting the corecivity ($H_c$) for all three samples. (h) Representative Raman data for all three samples depicting D-band at 1368 cm$^{-1}$, G band at 1580 cm$^{-1}$, and 2D at around 2700 cm$^{-1}$. $I_D/I_G$ for Fe@CNT (0.24) is one of the best reported so far.*

**Figure 4(g)** shows normalized $M_s$, limited to low field scans, for clear depiction of the coercively $H_c$ for each type of sample. Here the Ni@CNT exhibit very low coercivity, almost mimicking a superparamagnetic type of behavior. The $H_c$ for Fe and Co@CNT are ~ few 100 Oe and can be further tuned by variations in $T_{pyro}$, depending on the application specific requirement [26, 38-39]. For all the three samples, the Raman data to evaluate the crystallinity of the graphitic shells and the defects in $sp^2$-hybridized carbon structure is shown in **Figure 4(h)**. Raman data shows both the defect (D) and the graphitic (G) peaks at the position at 1358 cm$^{-1}$ and 1580 cm$^{-1}$ respectively. The third peak, corresponding to 2D or G` at ~ 2700 cm$^{-1}$, is also seen in all three samples, and the data is consistent with earlier reports on metal@CNT [1]. The ratio of intensities of defect and graphitic band, $I_D/I_G$, in case of Fe@CNT is 0.24 signifying well-formed graphitic shells in this case. This sample shows significant reduction in the catalytic particles, which is also evident in narrow and small Raman $I_D$ peak. In the case of Co@CNT, $I_D/I_G$ is ~ 0.6. In this case, some of the CoO$_x$ Raman peaks in the vicinity of the graphitic D peak also appear. The presence of relatively large density of residue metal particles outside the CNT, which are likely to get oxidized, can also contribute for oxide signal. In the case of Ni@CNT, $I_D/I_G$ obtained for the sample shown in **Figure 4(h)** is ~ 0.9, which can be significantly reduced by increasing the ratio of camphor to nickelocene, as explained in the next section.

It is to be noted that the morphology as well as the filling efficiency for all three type samples can be further controlled by variations in the synthesis parameters, including the ratio of camphor and metallocene. A few representatives are presented in the next section.

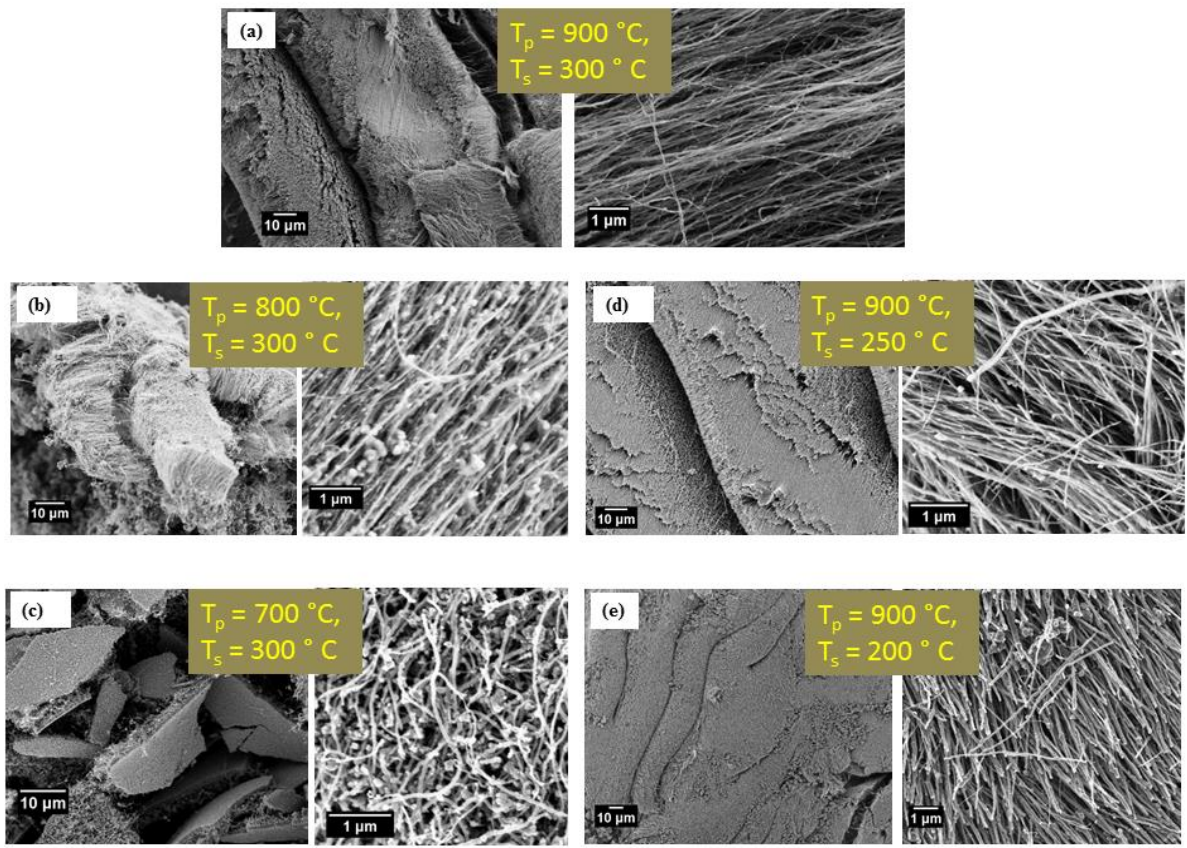

*Figure 5:* Representative SEM micrographs for Fe@CNT for **(a)** the best optimized parameters, $T_{pyro}$ = 900 °C and $T_{sub}$ = 300 °C. For a fixed $T_{sub}$, and Ar-Flow, change in $T_{pyro}$ results in tendency for entangled CNT and disrupted alignment and more residue particles, as shown in **(a, b, c)**. Images **(a, d, e)** show the variation in $T_{sub}$, while $T_{pyro}$ is held constant. Here the forest structures are retained for different $T_{sub}$ and same $T_{pyro}$, a slight increase in residual particle density is also observed for lower $T_{sub}$. All the samples shown in this figure have been collected from constant temperature region.

## 3.3: Variation in morphology with the change of synthesis parameters

### 3.3.1: Fe@CNT

The representative SEM images for CNT synthesized with varying $T_{pyro}$ and varying $T_{sub}$ is shown for Fe@CNT in **Figure 5**. As evident from the **Figures 5 (a-d-e)**, the aligned forests have been observed for keeping a fixed $T_{pyro} \sim$ 900 °C and varying $T_{sub}$. However, the residue particle density is slightly more for samples that are formed with lower $T_{sub}$. On keeping a fixed $T_{sub}$ but varying $T_{pyro}$, the aligned forest structure is disrupted. For instance spaghetti like entangles CNT are seen. This is evident from the **Figures 5 (a-b-c)**. The residue particle density is also seen to increase with decreasing $T_{pyro}$. These data enable us to optimize the best parameters for a significant filling efficiency along with minimum catalytic particles. Though there are reports on the variation of $T_{pyro}$, $T_{sub}$, and Ar-flow on the $M_s$ and $H_c$ of Fe@CNT [26, 38-39], the data presented here shows a systematic way to obtain aligned forest structures with significantly narrow L and D distribution, uniformity and systematic reduction in residue nano-particles adhering outside the CNT

*3.3.2: Co@CNT:*

**Figure 6** shows the morphology of the obtained CNT in which all synthesis parameters are similar to the Co@CNT shown in **Table 1**, except the rate of Ar-flow, which is lower in this case. Lowering the Ar-flow rate, no CNT are deposited in the pyrolysis zone. However, self-organized structures of Co@CNT are found with significant yield in variable temperature zone. **Figure 6** displays aligned forests of Co@CNT with a tendency to roll over. These tubular structures are fairly large, extending upto few 100 μ-m in length. The length of the CNT, in different rolled-over structures obtained from $T_{var}$ zone ranges from ~ 3–20 μ-m as can be seen in **Figures 6 (c)** and **(d)**. This again confirms the CNT deposition occurring in varying temperature region leads to significant variations in the length of the CNT. The overall data on Co@CNT suggests that manipulating the ratio of camphor and cobaltocene as a synthesis parameter, it is easy to form self-organized structures of Co@CNT. Various combination of all four synthesis parameters can be further tuned to obtain desired morphology and filling efficiency.

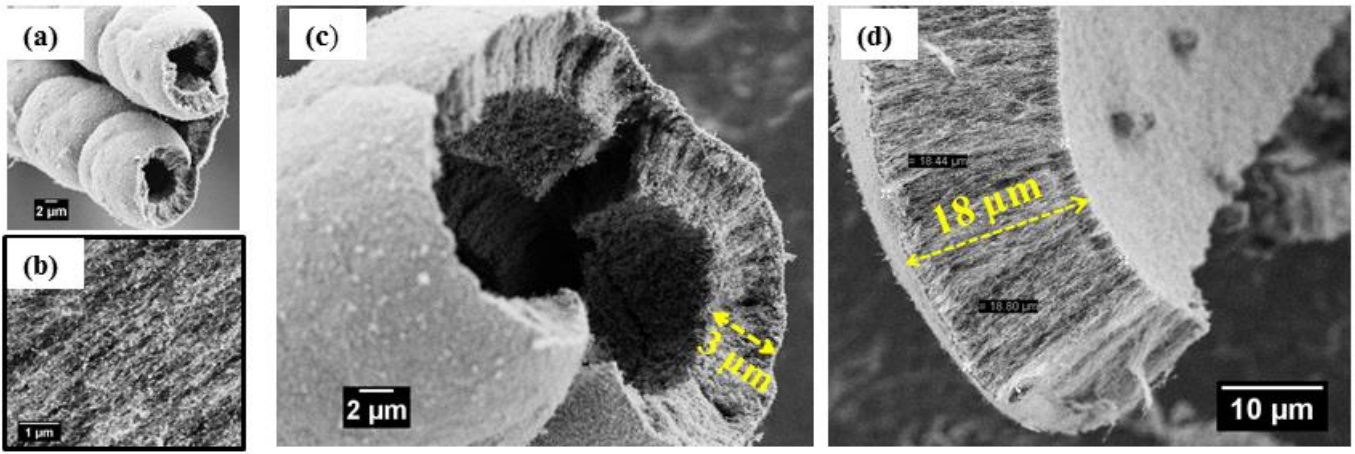

*Figure 6:* *Representative SEM micrographs of Co@CNT formed at $T_{pyro}$ = 800 °C and $T_{sub}$ = 200 °C. (a) displays fairly large rolled –over self organized structures obtained from variable temperature zone. The catalytic particle density is larger, as evident from (b). Images (c) and (d) indicate that Co@CNT obtained from the variable temperature zone exhibit a wider distribution in length of the CNT (3 – 20 µm) in these rolled –over structures.*

### 3.3.3: Ni@CNT:

This sample is relatively most difficult to obtain in the aligned forest morphology. However, the data presented here underlines the reasons that lead to the difficulties associated with the synthesis of Ni@CNT. Here we find a lot of inter correlation between the parameters, unlike the case of Fe@CNT, where the synthesis parameters are independent of each other and it is easy to optimize conditions for a higher filling fraction, narrow L and D distribution and minimum residue particle density outside the CNT. Similar to the case of Co@CNT, the Ni@CNT are formed only when camphor is mixed with nickelocene. From multiple runs, we conclude that a lower $T_{pyro}$ and higher $T_{sub}$, for a given Ar-flow, are more conducive to the formation of Ni@CNT. The outcome of the successful run is already shown in **Figure 2(f)**.

Attempts were made to obtain self-organized/aligned forest structure of Ni@CNT, similar to Co@CNT and Fe@CNT. For controlling the morphology for Ni@CNT, we varied camphor (C) to nickelocene (N). **Figure 7** shows the results of these trials. Keeping the ratio C: N as 1:5, the tendency to form aligned forest is observed in $T_{pyro}$ region as well as $T_{var}$ region, but with almost no evidence of CNT **Figure 7(a)**. Increasing the amount of camphor (C: N ~3:5) results in entangled *spaghetti*-like CNT in $T_{pyro}$ region, similar to what is shown in **Figure 2(f)**. For this run, however, the $T_{var}$ region displays some interesting sandwich type structure depicted in **Figure 7(b)**. Here micron size nano particle beds are formed and the CNT grow on either side of this nano particle bed, giving rise to sandwich type self-organized structures. These CNT which are formed on the either side of the nano-particle bed exhibit reduced $I_D/I_G$ ~ 0.6, signifying good quality graphitic shells. The $M_s$ value in bulk magnetization is likely to be much larger, due to the presence large number of catalytic particles in the form of nano carpet, but it will not represent the Ni filling within the graphitic shells, such as shown for Ni@CNT Figure **2(f)**. Increasing the amount of camphor (C:N ratio as 20:5) yields very long curled tubes, as shown in **Figure 7(c)**, and with further improved $I_D/I_G$ ~0.5. However, the $M_s$ value in this case is ~2-3 emu/g. It can be concluded that in the case of Ni@CNT, the slightest change in the parameters affects the morphology considerably, thus we conclude that synthesis parameters are correlated. Given the size of the synthesis chamber, the amount of nickelocene also plays a crucial role in deciding the alignment.

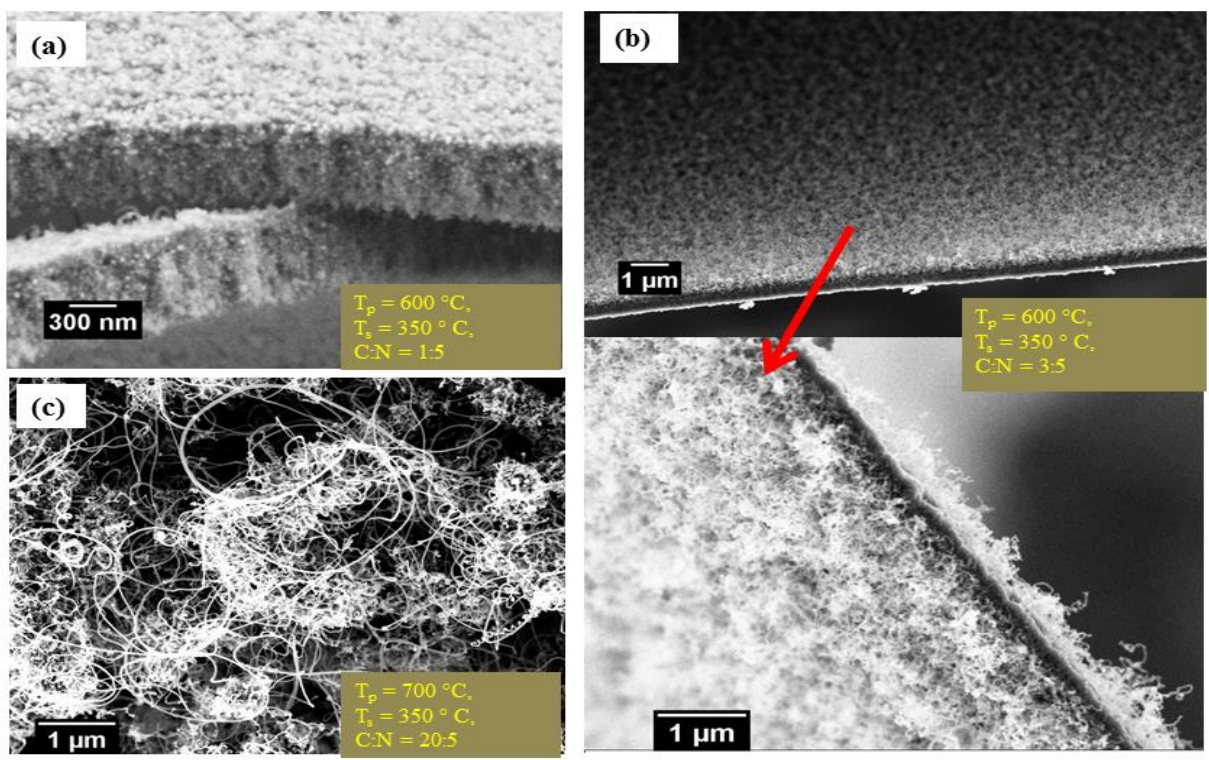

*Figure 7: Representative SEM micrographs of Ni@CNTs formed varying the amount of camphor (**a**) Aligned structures of metal particles are formed in $T_{pyro}$, region for camphor to nicklocene ratio as 1:5. (**b**) For camphor to nickelocene to 3:5, a bed of nanoparticles spread over microns in length is obtained, with CNT formed on the either side of the bed. Sample is obtained from $T_{var}$. In this case, $T_{pyro}$ region exhibits entangled CNT, similar to the ones shown in Figure 2f. (**c**) Long entangled tubes are obtained on increasing the camphor to nickelocene ratio to 20:5 at $T_{pyro}$ = 700 °C and $T_{sub}$ = 350 C.*

Overall, we conclude that any single zone furnace, which is routinely available, can be a better suited for the synthesis of CNT using pyrolysis of metallocene by careful temperature profiling. This is needed for identification of sublimation point and pyrolysis zone. A longer pyrolysis zone with better temperature accuracy is crucial for obtaining desired morphology with uniformity, narrow length and diameter distribution of the final product. Ar-gas flow and the sublimation temperature become more correlated, depending on the length of the variable temperature region between the sublimation point and the onset of the constant temperature zone of the furnace. For instance, Co and Ni@CNT samples are difficult to form by sublimation of their respective metallocene as only metal particles are observed along the length of the inner quartz tube, indicating that the carbon source is not sufficient. The addition of camphor, however, circumvents this problem and provides adequate source of carbon which facilitates the growth of metal filled CNT. This also adds another control parameter, the ratio of carbon and metallocene for tuning the morphology of filled CNT. This also appears to depend crucially on the exact dimensions of the synthesis chamber. There also exists a systematic pattern for obtaining self-organized structures in all three cases. The best optimized $T_{pyro}$ is highest for Fe@CNT (900 °C) and it is systematically reduced for Co@CNT (800 °C) and Ni@CNT (700 °C) as evident from **Table 1**.

## 4. Oxides@CNT

Once metal@CNT are formed with significant filling fraction, some post–synthesis annealing can produce oxides@CNT formed in the same morphology as metal@CNT [8]. The nature of application dictates the desired morphology of filled CNT. For instance, $Fe_2O_3$@CNT is relevant for both battery related as well as nano-electronics / spintronic applications [8, 10, 42] which would require spaghetti like morphology in batter applications whereas aligned forests can be more suited for nano scale device fabrication.

α- $Fe_2O_3$ in bulk form is a room temperature antiferromagnet that also exhibits the phenomenon of weak ferromagnetism at well-known Morin transition temperature ($T_M$) [10]. The weak ferromagnetism is known to be associated with the phenomenon of spin canting. α-$Fe_2O_3$ is also a symmetry allowed piezomagnet, wherein a magnetic moment can be generated by stress, given by the equation, $M_i = P_{ijk} \sigma_{jk}$, where σ is stress [43]. It is therefore interesting to encapsulate this compound inside CNT. This encapsulation is obtained by using as-prepared Fe@CNT, which is further annealed in $CO_2$ environment at 500 °C for 20 minutes. This heat treatment enables formation of Fe2O3@CNT, without disturbing the graphitic shells of the CNT. The representative SEM and TEM images of the representative of $Fe_2O_3$@CNT are shown in **Figure 8(a).** This sample has been characterized by lab XRD and match well with the graphitic peak corresponding to the CNT (ICSD code: 015840) and α-$Fe_2O_3$ peaks (ICSD code: 031170). This sample was also characterized by the synchrotron XRD data (Intensity vs 2-theta) as shown in **Figure 8(b)**. The solid black line black is fit to the experimental data (red dots) using Rietveld profile refinement. The refinement was done using two phase model with α-$Fe_2O_3$ and graphite with space group *R -3 c* and *P 63 m c* respectively. As evident from this fitting, both phases, α-$Fe_2O_3$ and graphite are identified and fitted and no other phase was detected in the synchrotron XRD data. The Rietveld refinement also reveals the percentage of α-$Fe_2O_3$ in the sample, which is ~ 65%, further confirming the high filling fraction. **Figure 8(c)** shows the magnetization as a function of temperature, clearly exhibiting the Morin transition, which signifies the onset of weak ferromagnetism and piezo magnetism in this CNT-metal oxide hybrid.

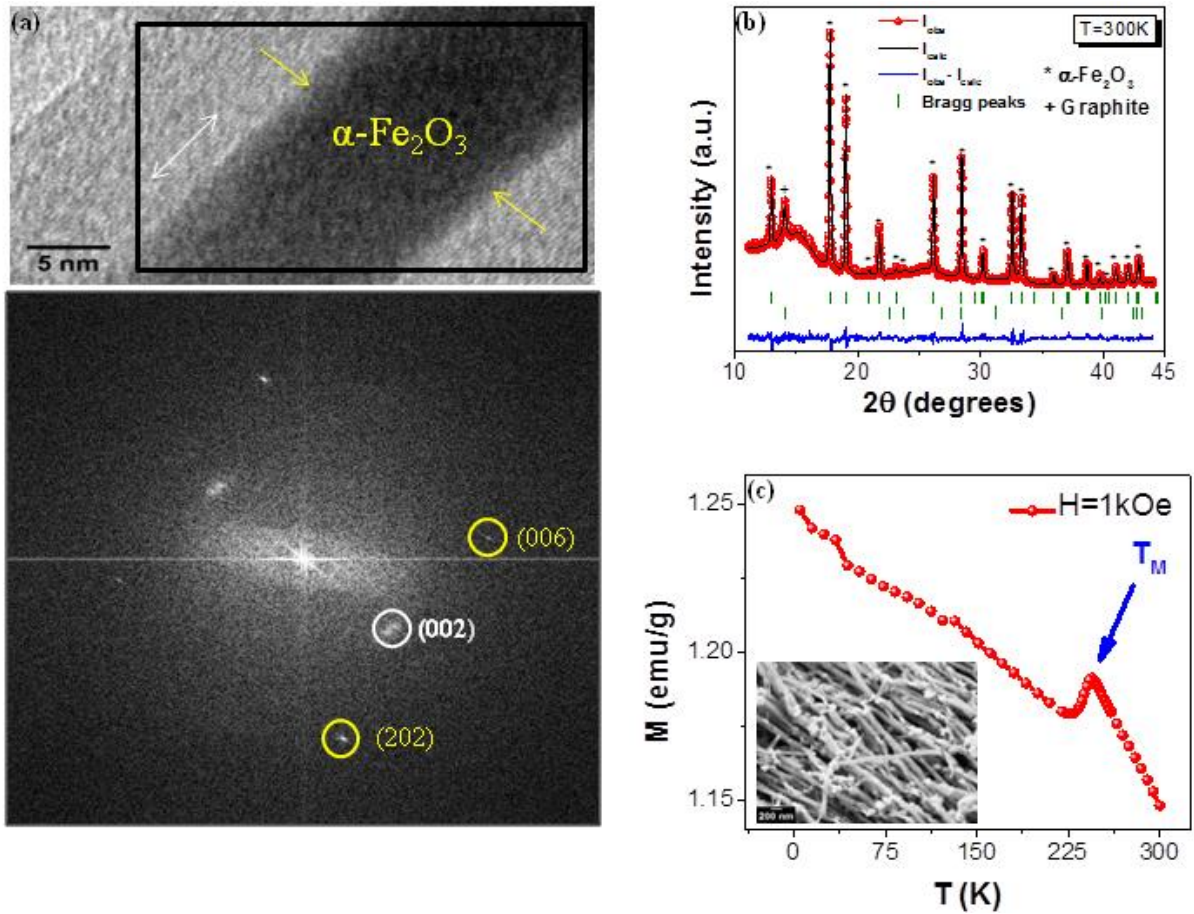

*Figure 8: (a) Representative TEM micrograph of Fe$_2$O$_3$@CNT with the corresponding Fast Fourier transform with yellow circles highlighting the reflections from the encapsulated α-Fe$_2$O$_3$. (002) denotes the graphitic planes. (b) Synchrotron XRD data of Fe$_2$O$_3$@CNT along with Rietveld Profile Refinement The goodness of fit obtained was 3.64. The red dots are the observed data, black line is best fit, and the blue line shows the difference in the observed and the calculated data. (c) Magnetization vs Temperature data at 1 kOe, depicting a magnetic transition at ~ 250 K. This is also known as the Morin Transition temperature ($T_M$), signifying the onset of weak ferromagnetism and piezo magnetism in an otherwise antiferromagnetic α-Fe$_2$O$_3$.*

## 5. Conclusion

We have synthesized Fe, Co and Ni@CNT using metallocene as a solid precursor following Chemical Vapor Deposition route. We demonstrate that a single zone furnace is sufficient to obtain high-quality multiwall CNT in which ferromagnetic metal is encapsulated efficiently. For the synthesis of Co and Ni@CNT, their respective metallocene are co pyrolyzed with a cost effective and environmental friendly compound camphor. Camphor provides an extra source of carbon and enables the formation Ni and Co@CNT with significantly higher filling efficiency than what is previously reported by using aerosol assisted CVD of metallocene. We show that co-pyrolysis of metallocene with camphor also leads to a variety of self organized structures that can be tailored by manipulating the ratio of metellocene to camphor. By systematic variation of synthesis parameters, it is possible to obtain self-organized structures of filled CNT, with narrow diameter and length distribution and reduced residue particle density. The metal@CNT samples can be obtained in desired morphology and can be used as a template to form oxide@CNT. These hybrids can have numerous applications in spintronic, magneto-optics and in the energy sector.

**Acknowledgements**

Authors thank Dr. Sunil Nair and Dr. Mukul Kabir (IISER, Pune) for SQUID magnetization measurements and useful contributions, Mr. J. Parmar and Mr. S.C. Purandare (TIFR) for TEM measurements, Mr. Vikram Bakaraju (IISER-P) for Raman measurements, DST and Saha Institute of Nuclear Physics, India, for facilitating the experiments at the Indian Beamline, Photon Factory, KEK, Japan. AB acknowledges Department of Science and Technology (DST), India for funding support through a Ramanujan Grant and the DST Nano mission Thematic Unit Grant (# SR/NM/TP-13/2016).